\documentclass[prd,superscriptaddress,amsfonts,amssymb,amsmath,showpacs,twocolumn]{revtex4-2}
\usepackage{bm}
\usepackage{amsfonts}
\usepackage{latexsym}
\usepackage[latin1]{inputenc}
\usepackage{graphicx}
\usepackage{amsmath}
\usepackage{palatino}
\usepackage{mathpazo}
\usepackage{textcomp}
\usepackage{float}
\usepackage{booktabs}
\usepackage{dcolumn}
\usepackage{ragged2e}
\usepackage{hyperref}
\hypersetup{colorlinks,citecolor=blue}
\hypersetup{colorlinks=true,linkcolor=red,filecolor=magenta,    urlcolor=blue}
\usepackage{amsmath}
\usepackage{xcolor}
\usepackage{orcidlink}
\usepackage[caption=false]{subfig}
\usepackage{commath}
\def\jnl@style{\it}
\def\aaref@jnl#1{{\jnl@style#1}}

\def\aaref@jnl#1{{\jnl@style#1}}

\def\aj{\aaref@jnl{AJ}}                   % Astronomical Journal
\def\apj{\aaref@jnl{ApJ}}                 % Astrophysical Journal
\def\apjl{\aaref@jnl{ApJ}}                % Astrophysical Journal, Letters
\def\apjs{\aaref@jnl{ApJS}}               % Astrophysical Journal, Supplement
\def\apss{\aaref@jnl{Ap\&SS}}             % Astrophysics and Space Science
\def\aap{\aaref@jnl{A\&A}}                % Astronomy and Astrophysics
\def\aapr{\aaref@jnl{A\&A~Rev.}}          % Astronomy and Astrophysics Reviews
\def\aaps{\aaref@jnl{A\&AS}}              % Astronomy and Astrophysics, Supplement
\def\mnras{\aaref@jnl{Mon.~Not.~Roy.~Astron.~Soc.}}             % Monthly Notices of the RAS
\def\prd{\aaref@jnl{Phys.~Rev.~D}}        % Physical Review D
\def\plb{\aaref@jnl{Phys.~Lett.~B}}        % Physics Letters B
\def\prc{\aaref@jnl{Phys.~Rev.~C}}  % Physical Review C
\def\prl{\aaref@jnl{Phys.~Rev.~Lett.}}    % Physical Review Letters
\def\qjras{\aaref@jnl{QJRAS}}             % Quarterly Journal of the RAS
\def\skytel{\aaref@jnl{S\&T}}             % Sky and Telescope
\def\ssr{\aaref@jnl{Space~Sci.~Rev.}}     % Space Science Reviews
\def\zap{\aaref@jnl{ZAp}}                 % Zeitschrift fuer Astrophysik
\def\nat{\aaref@jnl{Nature}}              % Nature
\def\aplett{\aaref@jnl{Astrophys.~Lett.}} % Astrophysics Letters
\def\apspr{\aaref@jnl{Astrophys.~Space~Phys.~Res.}} % Astrophysics Space Physics Research
\def\physrep{\aaref@jnl{Phys.~Rep.}}      % Physics Reports
\def\physscr{\aaref@jnl{Phys.~Scr}}       % Physica Scripta
\def\commat{\aaref@jnl{Comm.~Math.~Phys.}}              % Communications in Mathematical Physics
\def\science{\aaref@jnl{Science}}               % Science
\def\cqg{\aaref@jnl{Classical Quant.~Grav.}}            % Classical and Quantum Gravity
\def\jpcs{\aaref@jnl{JPCS}}                                     % Journal of Physics Conference Series
\def\ijmpd{\aaref@jnl{Int.~J.~Mod.~Phys.~D}}                    % International Journal of Modern Physics D
\def\grg{\aaref@jnl{Gen.~Relat.~Gravit.}}               % General Relativity and Gravitation
\def\rpp{\aaref@jnl{Rep.~Prog.~Phys.}}          % Reports on Progress in Physics
\def\npa{\aaref@jnl{Nucl.~Phys.~A}}        % Nuclear Physics A
\def\lrr{\aaref@jnl{Living Rev.~Rel.}}                   % Living reviews in relativity
\def\jcap{\aaref@jnl{J.~Cosmology Astropart.~Phys.}}    % Journal of cosmology and astroparticle physics
\def\rmp{\aaref@jnl{Rev.~Mod.~Phys.}}   %Reviews of modern physics
\def\epjc{\aaref@jnl{Eur.~Phys.~J.~C}}

\allowdisplaybreaks[1]
\begin{document}
%\color{red}
\color{black}       %% For one column
\title{Anisotropic quark stars in $R^2$ gravity}

\author{Grigoris Panotopoulos \orcidlink{0000-0002-7647-4072}} \email{grigorios.panotopoulos@tecnico.ulisboa.pt}
\affiliation{Centro de Astrof{\'i}sica e Gravita{\c c}{\~a}o-CENTRA, Instituto Superior T{\'e}cnico-IST, Universidade de Lisboa-UL, Av. Rovisco Pais, 1049-001 Lisboa, Portugal}

\author{Takol Tangphati
\orcidlink{0000-0002-6818-8404}} 
\email{takoltang@gmail.com}
\affiliation{Department of Physics, Faculty of Science, Chulalongkorn University, \\Bangkok 10330, Thailand}

\author{Ayan Banerjee \orcidlink{0000-0003-3422-8233}} 
\email{ayanbanerjeemath@gmail.com}
\affiliation{Astrophysics and Cosmology Research Unit, School of Mathematics, Statistics and Computer Science, University of KwaZulu--Natal, Private Bag X54001, Durban 4000, South Africa}

 \author{M. K. Jasim}
 \email{mahmoodkhalid@unizwa.edu.om}
\affiliation {Department of Mathematical and Physical Sciences, College of Arts and Sciences,\\ University of Nizwa, Nizwa, Sultanate of Oman}

%%%%%%%%%%%%%%%%%%%%%%%%%%%%%%%%%%%%%  DATE  %%%%%%%%%%%%%%%%%%%%%%%%%%%%%%%%%%%%

\date{\today}

\begin{abstract}
In the present paper, we treat the problem of the existence of quark stars (QSs) for selected homogeneous and unpaired charge-neutral $3$-flavor interacting quark matter with $\mathcal{O}(m_s^4)$ corrections equations of state (EoS). Using the EoS combined with the Tolmann-Oppenheimer-Volkoff (TOV) structure equations, the properties of stars are explored by obtaining their mass-radius relations. All calculations are carried out within the framework of the $R$-squared gravity defined by $f(R) = R+a R^2$. Our main goal is to discuss the effect for a wide range of the $R$-squared gravity parameter, $a$, on the mass-radius and the mass-central mass density $(M-\epsilon_c)$ relation of QSs. Furthermore, we investigate the dynamical stability condition for those stars, and we show that their dynamical stability depends on the anisotropic parameters $ B_{\perp}$ and $a_4^{\perp}$ coming from the respective EoS. In such a scenario our results provide circumstantial evidence in favor of super-massive pulsars.
\end{abstract}

\maketitle

\section{Introduction}

Numerous modified gravity theories have been proposed to address several shortcomings coming out in the study of gravitational interaction at infra-red and ultra-violet scales. There are also hints that the structure of gravity may be different in the infra-red. Moreover, the current accelerating expansion of the universe is one of the biggest unsolved problems in fundamental physics which can be hardly explained through the most successful general relativity (GR) theory. For this reason, scientists have come up with alternative gravity theories extending Einstein's theory at the level of an effective action. In this perspective, it is possible to obtain accelerating expansion of the universe without assuming the so-called \textit{dark energy} as a new material field. Therefore, in recent years, theory of modified gravity has an active research area, and in this article we wish to pursue this approach.

Amongst several different modified gravity theories, $f(R)$-gravity, where $f(R)$ is a generic function of the Ricci scalar, $R$, has been extensively explored in the literature \cite{Sotiriou:2008rp,DeFelice:2010aj,Nojiri:2010wj,Nojiri:2017ncd}. This modification comes into the game by replacing the Einstein-Hilbert Lagrangian, which is linear in the Ricci scalar, with a more general function of the curvature $f(R)$.  In particular, $f(R)$ modifications of GR give rise to phenomenologically viable cosmological models, both for the early and late time acceleration of the Universe, instead of searching for new dynamical components \cite{Clifton:2011jh,Capozziello:2011et}. Furthermore,
$f(R)$ gravity which reduces to GR for the case of $f(R) \to R$.
In this sense, although many models with different 
functions of curvature terms have been proposed, Starobinsky's model $f(R)=R+aR^2$ \cite{Starobinsky:1980te} is perhaps the most relevant one and favored by CMB data \cite{Akrami:2018odb}. Moreover, it provides us with a natural inflationary era in the early universe, and it does not contain ghost-like modes.

The strong gravity regime for the construction of compact objects is yet another way to test the viability of alternative theories of gravity. Therefore, the astrophysical implications of the Starobinsky model should be investigated as well. In this context, massive and compact neutron stars whose mass is around $\sim 2 M_{\odot}$ have been studied (see, e.g., Refs. \cite{Yazadjiev:2015zia,Astashenok:2017dpo,Blazquez-Salcedo:2018qyy} and references therein). Recent successful probes of a compact binary merger with a 22.2 - 24.3 $M_{\odot}$ black hole and a 
compact object with mass 2.50 - 2.67 $M_{\odot}$ have also taken place
in the new era of gravitational wave (GW) observational Astronomy from GW 190412 \cite{LIGOScientific:2020stg} 
and  GW 190814 \cite{Abbott:2020khf} by the aLIGO Scientific and Virgo Collaboration. 
The event GW 190814 is a challenging one, since it may be either the most massive neutron star (NS) or the lightest black hole ever observed. Thus, the gravitational wave signal GW 190814 has started a discussion on the nature of the secondary object with a mass of $2.6~M_{\odot}$. One of the proposals that have been put forward is modified gravity \cite{Astashenok:2020qds,Astashenok:2021peo}.

Furthermore, in \cite{Panotopoulos:2018enj} the authors have studied non-rotating dark stars. They assumed  dark matter to be bosonic and self-interacting modelled inside the star. Correspondingly, the structure of compact stars in perturbative $f(R)$ gravity was studied in 
\cite{Arapoglu:2010rz,Alavirad:2013paa,Astashenok:2013vza,Astashenok:2014pua}, according to which the scalar curvature $R$ is defined by Einstein equations at zeroth order on the small parameter,
i.e. $R \sim T$, where $T$ is the trace of the energy-momentum tensor. Moreover, non-perturbative approaches are also available in Ref. \cite{Capozziello:2015yza}. The causal maximum mass limit of NSs in $f(R)$ gravity focusing on the $R^2$ model has been studied in \cite{Astashenok:2021peo}.

In this connection, it may be of some interest to study the structure of quark stars in the Starobinsky model of gravity. For the quark matter equation of state (EoS)  we consider homogeneously confined matter inside the star with  3-flavour neutral charge and a fixed strange quark mass \cite{Flores:2017kte}. Moreover, recent instrumentation and computational advances speculate that interactions among quarks may generate changes at the interior of the shell generating anisotropies i.e., the interior pressure in the radial direction being different from that in the polar or azimuthal directions. In fact, pressure anisotropy affects the mass-radius relation and the gravitational red-shift of compact stars. The idea of an anisotropic relativistic sphere was first pointed out by Lema\'ite \cite{Lema}. After that Bowers and Liang \cite{bowers} boosted the whole idea considering the pressure anisotropy on neutron stars (NSs). However, the most precious observation came from Ruderman \cite{Ruderman}, who observed that nuclear matter tends 
to become anisotropic at very high densities of order $10^{15}$  g/cm$^3$. Moreover, in \cite{Herrera:1997plx} the authors have shown that anisotropic stars could exist in a strong gravity region. 

The plan of the paper is as follows: In Section \ref{sec2} we formulate the problem and the reduced field equations that will be solved numerically for the particular case of Starobinsky's model. We present the TOV equations for static, spherically symmetric stellar metric. In Section \ref{sec3} we present an overview of a QCD motivated EoS. In section \ref{sec4} we present and discuss the results for QSs with homogeneously confined matter inside the star in $R$-squared gravity. In this scenario, we show the possibility of obtaining the maximal stellar mass which satisfies the recent observational data for PSR J1614-2230. We discuss the final outcomes and future perspectives in Section \ref{sec5}.  Here we adopt the signature $(-, +, +, +)$, and set $c = 1$.

\section{Field Equations  and set up}\label{sec2}

In this section we describe the analytical setup of the problem, and we and present the equations of motion. The action of $f(R)$ theories is given by
\begin{eqnarray}\label{A}
S= \frac{1}{16\pi G} \int d^4x \sqrt{-g} f(R) + S_{\rm
matter}(\psi_i, g_{\mu\nu}),
\end{eqnarray}
where $G$ is Newton's constant and $R$ is the scalar curvature with respect to the spacetime metric $g_{\mu\nu}$. Since, $S_{\rm matter}$ is the action of matter field depending on the metric tensor and the matter fields $\psi_i$. To avoid pathological situations, such as ghosts and tachyonic instabilities, the 
viable $f(R)$ theories have to satisfy the following conditions \cite{Sotiriou:2008rp,DeFelice:2010aj}
\begin{eqnarray}
\frac{d^2f}{dR^2}\ge 0,  \;\;\; \frac{df}{dR}>0,
\end{eqnarray}
respectively.

In the discussion to follow we shall consider the specific form of Starobinsky's model, i.e. $f(R)=R+aR^2$, where the free parameter $a$ satisfying $a \geq 0$ in agreement with the above mentioned inequalities \cite{Arapoglu:2010rz}. The special value $a = 0$ corresponds to GR. Since $a$ has dimensions of $[mass]^{-2}$.  
we may write it in the form $a = 1/M^2$, where now the mass scale $M$ is the free parameter of the theory.

Tackling the problem corresponding to fourth order equations can be mathematically troublesome, metric $f(R)$ theories however can be recast as a scalar-tensor theory with only second order equations under a conformal transformation. Thus, we define the Einstein frame by performing the conformal transformation \cite{Brax:2008hh,Woodard:2006nt,Staykov:2014mwa}
\begin{equation}
\tilde{g}_{\mu \nu} = p g_{\mu \nu} = A^{-2} g_{\mu \nu},
\end{equation}
where $A (\phi) = \exp(-\phi/\sqrt{3})$. Now, the equation (\ref{A}) is equivalent to the more familiar equation
\begin{eqnarray}\label{act}
&& S = \frac{1}{16 \pi G} \int d^4x \sqrt{-\tilde{g}} [\tilde{R}-2 \tilde{g}^{\mu \nu} \partial_\mu \phi \partial_\nu \phi-V(\phi)] \nonumber \\
&&+ S_M[\psi_i, \tilde{g}_{\mu \nu} A(\phi)^2],
\end{eqnarray}
where the potential $V(\phi)$ is given by (see Ref. \cite{Brax:2008hh,Woodard:2006nt} for details) 
\begin{equation}
V(\phi) = \frac{(p-1)^2}{4 a p^2} = \frac{(1-exp(-2 \phi/\sqrt{3}))^2}{4 a}.
\end{equation}

Now, varying the equation (\ref{act}) with respect to $g_{\mu \nu}$ and $\phi$, we obtain Einstein's field equations as well as the Klein-Gordon equation. The field equations then become
\begin{eqnarray}
&& \tilde{G}_{\mu \nu} = 8 \pi G [ \tilde{T}_{\mu \nu} + T^{\phi}_{\mu \nu} ], \\
&& \nabla_\mu \nabla^\mu{\phi}-\frac{1}{4} V_{,\phi} = -4 \pi G \alpha \tilde{T},
\end{eqnarray}
where $T^{\phi}_{\mu \nu}$ is the stress-energy tensor corresponding to the scalar field and $V_{,\phi} \equiv \frac{dV(\phi)}{d\phi}$.
We note from Ref. \cite{Staykov:2014mwa,Yazadjiev:2014cza} that, there exists a direct coupling between matter and the scalar field with the coupling constant being $\alpha=-1/\sqrt{3}$. The Einstein frame energy-momentum tensor $\tilde{T}_{\mu \nu}$ is related to the one in the Jordan frame $T_{\mu \nu}$ via $\tilde{T}_{\mu \nu} = A(\phi)^2 T_{\mu \nu}$. In the following, the standard matter source under consideration is described as  an anisotropic fluid, for which the energy-momentum tensor is
\begin{eqnarray}
T^\nu_i &=& (\epsilon+ P)u^\nu u_i + P_{\perp} g^\nu_i + (P-P_{\perp})\chi_i \chi^\nu, 
\end{eqnarray}
where $\epsilon$ is the fluid energy density. Then the energy-momentum tensor contains only the following nonzero diagonal components: $T^\nu_i = \left( -\epsilon, P, P_{\perp}, P_{\perp} \right) $.

Here,  the energy density, the pressure components (radial and tangential) and the 4-velocity in the two frames are related via the formulae \cite{Yazadjiev:2014cza}
\begin{eqnarray}
\tilde{\epsilon} & = & A(\phi)^4 \epsilon \\
\tilde{P} & = & A(\phi)^4 P \\
\tilde{P}_{\perp} & = & A(\phi)^4 P_{\perp} 
\end{eqnarray}
where the tilde indicates the Einstein frame.

Since the purpose of the present paper is to study the structure of
anisotropic quark stars without rotation in $f(R)$ gravity, let us assume a static, spherically symmetric metric tensor describing interior solutions of the star
\begin{eqnarray}
ds^2= - e^{2\Phi(r)}dt^2 + e^{2\Lambda(r)}dr^2 + r^2 d \Omega^2,
\end{eqnarray}
where $d \Omega^2 = d\theta^2 + \sin^2\theta d\vartheta^2 $ is the line element on the unit 2-sphere. Moreover, the metric functions $\Phi(r)$ and $\Lambda(r)$ depend on the radial coordinate $r$, respectively. 

With those conditions imposed, the non-zero components of the field equations are the following
\begin{widetext}
\begin{eqnarray}
&&\frac{1}{r^2}\frac{d}{dr}\left[r(1- e^{-2\Lambda})\right]= 8\pi G
A^4(\phi) \epsilon + e^{-2\Lambda}\left(\frac{d\phi}{dr}\right)^2
+ \frac{1}{2} V(\phi), \label{eq:FieldEq1} \\
&&\frac{2}{r}e^{-2\Lambda} \frac{d\Phi}{dr} - \frac{1}{r^2}(1-
e^{-2\Lambda})= 8\pi G A^4(\phi) P +
e^{-2\Lambda}\left(\frac{d\phi}{dr}\right)^2 - \frac{1}{2}
V(\phi),\label{eq:FieldEq2}\\
&&\frac{d^2\phi}{dr^2} + \left(\frac{d\Phi}{dr} -
\frac{d\Lambda}{dr} + \frac{2}{r} \right)\frac{d\phi}{dr}= 4\pi G
\alpha(\phi)A^4(\phi)(\epsilon-P -2 P_{\perp})e^{2\Lambda} + \frac{1}{4}
\frac{dV(\phi)}{d\phi} e^{2\Lambda}, \label{eq:FieldEq3}\\
&&\frac{dp}{dr}= - (\epsilon + p) \left(\frac{d\Phi}{dr} +
\alpha(\phi)\frac{d\phi}{dr} \right)+{2 \over r}\left(P_{\perp} - P\right), \label{eq:FieldEq4}
\end{eqnarray}
\end{widetext}
The above formulation is used to study the interior structure of QSs, i.e. the two metric potentials, the energy density, two components of pressure and the scalar field. The above equations comprise the modified TOV equations, and they are clearly reduced to the standard GR
TOV equations when the scalar field is absent.

The first stage of our work is to numerically solve the modified TOV equations (\ref{eq:FieldEq1})-(\ref{eq:FieldEq4})
inside and outside the star simultaneously with the following natural Einstein frame boundary conditions at the center of the star
\begin{eqnarray}
\epsilon(0) = \epsilon_{c},\;\;\; \Lambda(0)=0,\;\;\;
\phi(0) = \phi_c, \;\;\; \frac{d\phi}{dr}(0)=0,\label{eq:BC1}
\end{eqnarray}
where $\epsilon_c$ and $\phi_c$ are the central values of the energy density and of the scalar field, respectively, while at infinity
\begin{eqnarray}\label{BCINF}
\lim_{r\to \infty}\Phi(r) = 0, \;\;\;\; \lim_{r\to \infty}\phi
(r)=0, \label{eq:BC2}
\end{eqnarray}
and the radius of the star $r_S$ is identified by the 
condition
\begin{eqnarray}\label{NSR}
P(r_S)=0.
\end{eqnarray}

At this point a comment on the boundary conditions is in order. The regularity of the scalar field $\phi$ is ensured by the condition $\frac{d\phi}{dr}(0) = 0$ at the center of the star, $r = 0$. 
To guarantee regularity at the origin as of the Einstein frame, we require $\Lambda(0)=0$.  Since the Einstein and the Jordan frame metrics are related via nonsingular conformal factor which ensures the regularity of the Jordan frame geometry at the center of the star also. The boundary conditions must be chosen so that they satisfy the asymptotic flatness requirement at infinity. This requires $\lim_{r\to\infty} V(\phi(r)) = 0$, which implies
$\lim_{r\to\infty}\phi(r) = 0$. Note that the conditions (\ref{BCINF})
ensure the asymptotic flatness in both Einstein and Jordan frame.

The radius $r_S$ of the star is defined as the distance to the center where the radial pressure vanishes, and the true radius of the star, as measured in the physical Jordan frame, is given by
\begin{eqnarray}
R_{S}= A[\phi(r_S)] r_S.
\end{eqnarray}

Finally, we are in a position to solve the TOV equations (\ref{eq:FieldEq1})-(\ref{eq:FieldEq4})
for a given EoS $P = P(\epsilon)$ that relates the pressure with the energy density with the following initial conditions. Next, we present the structure equations describing hydrostatic equilibrium of a QCD motivated EoS, as a matter source for QSs in the Starobinsky model.

%%%%%%%%%%%%%%%%%%%%%%%%%%%%%%FIGURES%%%%%%%%%%%%%%%%%%%%%%%%%%%%

\begin{figure}[h]
    \centering
    \includegraphics[width = 8 cm]{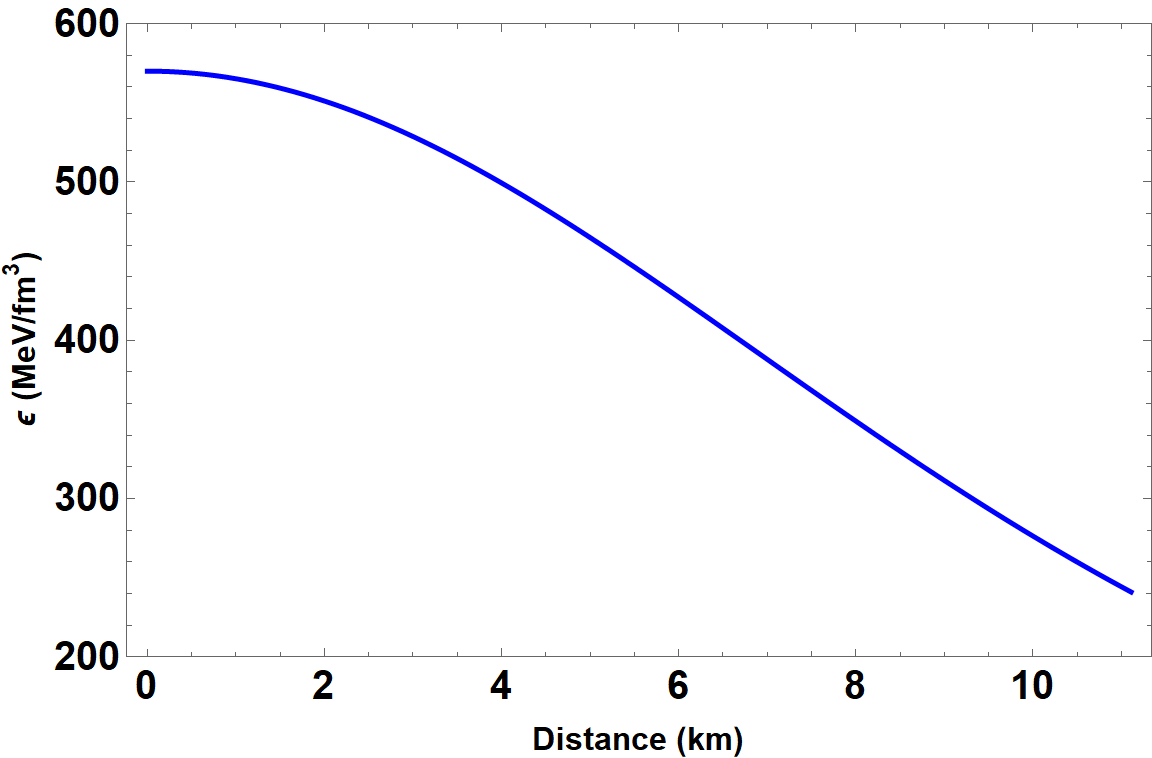}
    \caption{Variation of the energy density from the center of the star.}
    \label{Density}
\end{figure}

%%%%%%%%%%%%%%%%%%%%%

\begin{figure}[h]
    \centering
    \includegraphics[width = 8 cm]{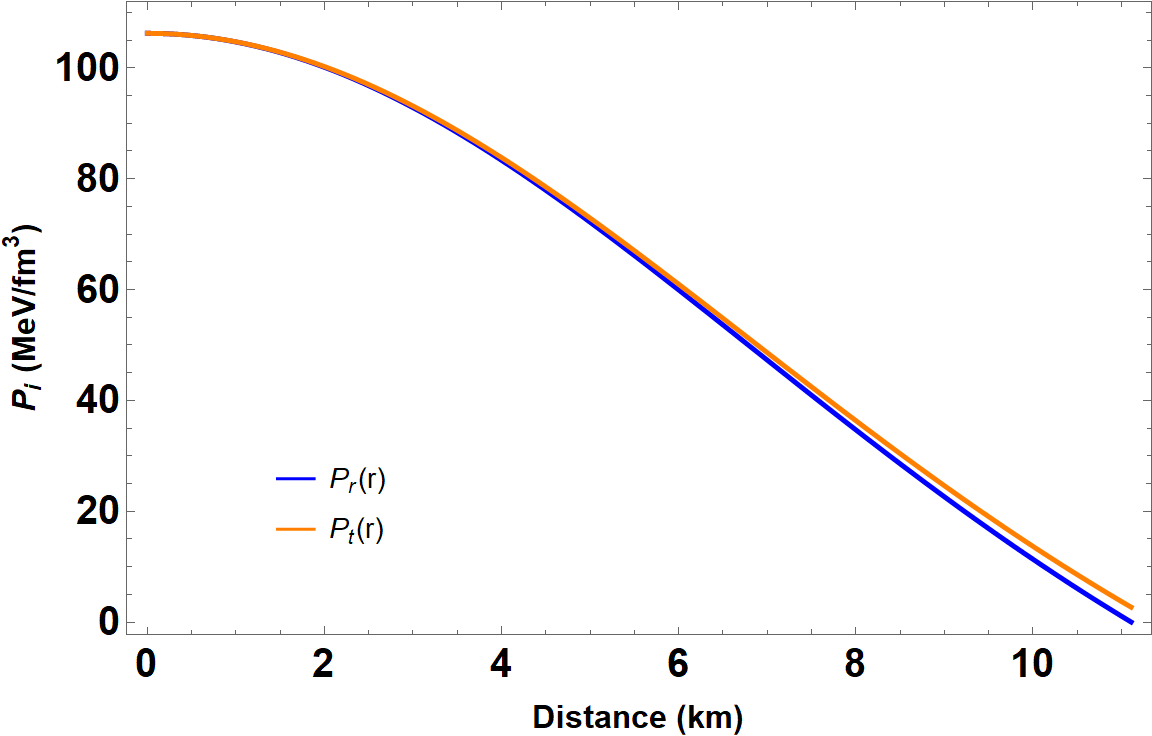}
    \caption{The pressure components in radial and tangential directions in the stellar interior.}
    \label{Pressure}
\end{figure}

\begin{figure}[h]
    \centering
    \includegraphics[width = 8 cm]{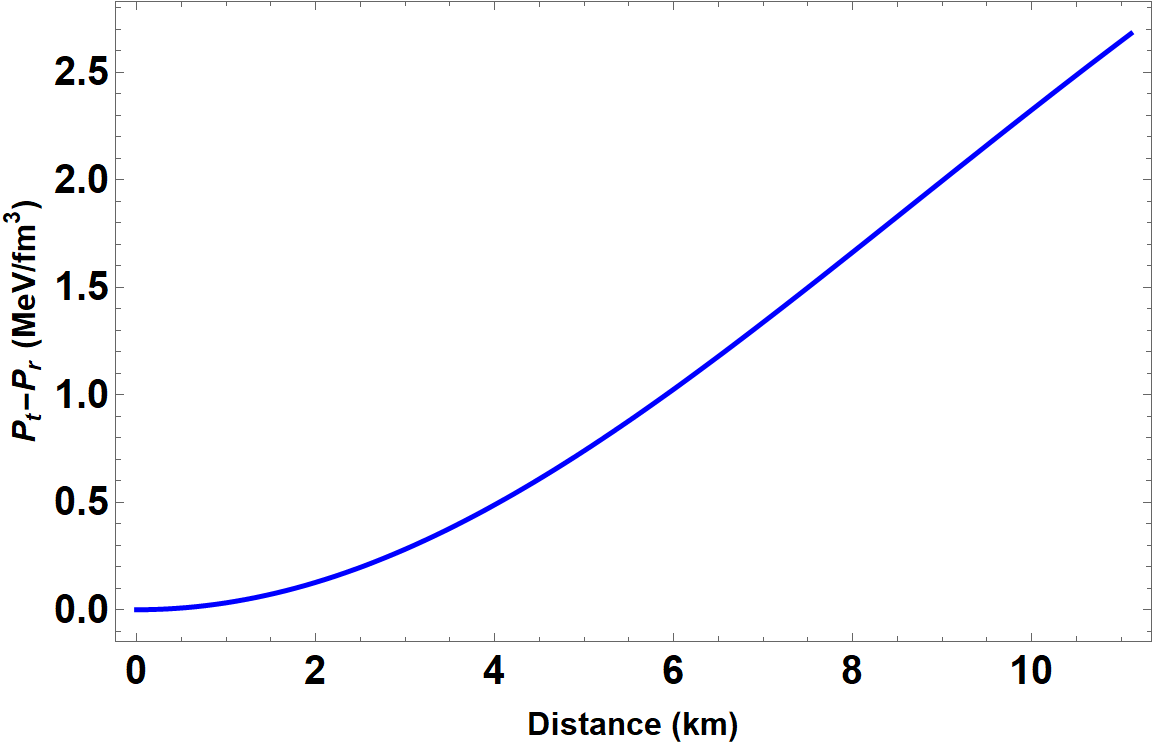}
    \caption{The plot depicts the pressure difference $\Delta = P_{\perp} - P$, where one verifies $\Delta > 0$.}
    \label{Ptessure}
\end{figure}

%%%%%%%%%%%%%%%%%%%%%%%%

%%%%%%%%%%%%%%%%%%%%%%%%%%%%%%%%%%%%%%%%%%%%%%%%%%%%%%%%%%%%
\section{Equations of State for Quark matter } \label{sec3}
%%%%%%%%%%%%%%%%%%%%%%%%%%%%%%%%%%%%%%%%%%%%%%%%%%%%%%%%%%%%

In this section we present the EoS that we will be using in modeling the strange star made of interacting quark matter, since it is believed that high density and relatively low temperature required to produce color superconducting quark matter. It is well known that superconductivity is a general feature of degenerate Fermi systems,
such as electrons, Helium 3 and nucleon systems. At sufficiently high baryon densities, when nucleons are converted to quark matter, the resulting quark matter composed of an infinite number of quarks being in a superconducting state at low enough temperatures  \cite{Alford:2001dt,Anglani:2006br}. For the degenerate Fermi gas, the attractive quark interaction makes the Fermi surface unstable. As recognized by Bardeen, Cooper and Schrieffer (BCS) this instability leads to the formation of the condensate of quark pairs (diquarks) with
nontrivial color structure. Detailed discussions about numerical calculations of color superconducting gaps were firstly
carried out by Bailin and Love \cite{Bailin:1983bm}. Thus, color superconductivity in quark matter \cite{Blaschke}
has become a compelling topic in quantum chromodynamics (QCD) over the years.

In QCD the quark-quark (qq) interaction not only is strong and attractive in many channels, but also many degrees of freedom are possible. This possibility leads to the different color 
superconducting phases \cite{Alford:2002kj,Alford:2005wj,Mannarelli:2007bs} depending on color and flavor as well as spin degrees of freedom.  In this asymptotic region the most favored state is the so called color-flavor-locked (CFL) phase \cite{Alford2002,Steiner2002}, in which quarks of all three colors and all three flavors form Cooper pairs \cite{Alford:1998mk}. However, there is strong reason to believe that these are not the only pairing patterns that are relevant in nature, and this open up the possibility of perturbative corrections in QCD. 

To set the stage for our analysis, we consider the EoS that consists of homogeneous and unpaired, overall electrically neutral, 3-flavor interacting quark matter \cite{Flores:2017kte}. But for simplicity, we describe this phase using the simple thermodynamic Bag model EoS \cite{Alford:2004pf} with $\mathcal{O}$ $(m_s^4)$ corrections.  In this set up several authors predict the existence of QSs stars made up of interacting quark EoS at ultra-high densities
\cite{Becerra-Vergara:2019uzm} (see for more \cite{Banerjee:2020dad}). Hence, 
the interacting quark EoS reads \cite{Becerra-Vergara:2019uzm}

\smallskip

\begin{widetext}
\begin{equation} \label{Prad1}
P=\dfrac{1}{3}\left(\epsilon-4B\right)-\dfrac{m_{s}^{2}}{3\pi}\sqrt{\dfrac{\epsilon-B}{a_4}}
+\dfrac{m_{s}^{4}}{12\pi^{2}}\left[1-\dfrac{1}{a_4}+3\ln\left(\dfrac{8\pi}{3m_{s}^{2}}\sqrt{\dfrac{\epsilon-B}{a_4}}\right)\right],
\end{equation}
\end{widetext}
where $\epsilon$ is the energy density of homogeneously distributed quark matter (also to $\mathcal{O}$ $(m_s^4)$ in the Bag model). Generally accepted values of Bag constant $B$ lies in the range of  $57\leq B \leq 92$ MeV/fm$^3$ \cite{Burgio:2018mcr,Blaschke:2018mqw}, whereas the strange quark mass is $m_{s}$ to be $100 \,{\rm MeV}$ \cite{Beringer:2012} for  our entire calculations. Another parameter $a_4$ comes from the QCD corrections on the pressure of the quark-free Fermi sea; is more relevant than the bag model parameter for  $M-R$ relation of QSs.

\smallskip

In the following, we consider another EoS for the tangential pressure that better describes an anisotropic QS, as follows  \cite{Becerra-Vergara:2019uzm}:
\begin{widetext}
\begin{eqnarray} \label{Prad2}
P_{\perp}= P_c +\dfrac{1}{3}\left(\epsilon-4B_{\perp}\right)-\dfrac{m_{s}^{2}}{3\pi}\sqrt{\dfrac{\epsilon-B_{\perp}}{a_4^{\perp}}}
+\dfrac{m_{s}^{4}}{12\pi^{2}}\left[1-\dfrac{1}{a_4^{\perp}}+3\ln\left(\dfrac{8\pi}{3m_{s}^{2}}\sqrt{\dfrac{\epsilon-B_{\perp}}{a_4^{\perp}}}\right)\right] \nonumber\\
-\dfrac{1}{3}\left(\epsilon_c-4B_{\perp}\right)+\dfrac{m_{s}^{2}}{3\pi}\sqrt{\dfrac{\epsilon_c-B_{\perp}}{a_4^{\perp}}}
-\dfrac{m_{s}^{4}}{12\pi^{2}}\left[1-\dfrac{1}{a_4^{\perp}}+3\ln\left(\dfrac{8\pi}{3m_{s}^{2}}\sqrt{\dfrac{\epsilon_c-B_{\perp}}{a_4^{\perp}}}\right)\right],
\end{eqnarray}
\end{widetext}
where $\epsilon_c$ and $P_c$ are the central energy density and central radial pressure for Eq. (\ref{Prad2}).  Our results for the energy density and pressure components, arising from a numerical elaboration, are presented Figs. \ref{Density} and \ref{Pressure}, respectively. 
In these figures we have chosen the parameter values $a = 5 \text{ km}^2$,  $B = 57 \text{ MeV/fm}^3$, $B_{\perp} = 70 \text{ MeV/fm}^3$, $a_4 = 0.9$, $a_{4} ^\perp = 0.3$ and
$\epsilon_{c} = 10 B$, respectively. Fig. \ref{Ptessure} shows that the radial and tangential pressures  are the same for $B = B_{\perp}$ and  $a_4 = a_4^{\perp}$  at the center of the star. This condition represents the case of an isotropic fluid at the stellar center $r = 0$. In the bottom panel of Fig. \ref{Ptessure} we depict $\Delta = P_{\perp} - P$, where one verifies $\Delta > 0$. Note that $\Delta = 0$ at the origin, $r = 0$, as was to be expected.

%%%%%%%%%%%%%%%%%%%%%%%%BEGINNING-OF-FIGURES%%%%%%%%%%%%%%%%%%%%%%

\begin{figure}[h]
    \centering
    \includegraphics[width = 8 cm]{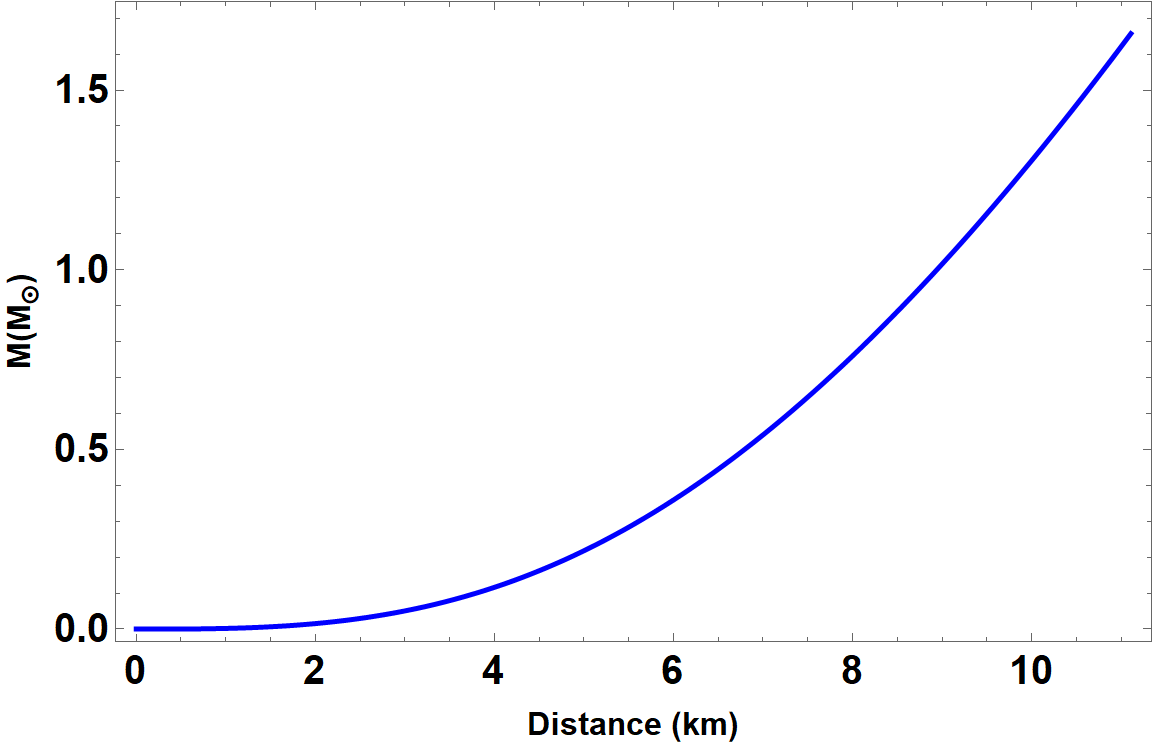}
    \caption{The graphs of the accumulated mass from the center.}
    \label{Mass}
\end{figure}

%%%%%%%%%%%%%%%%%%

\begin{figure}
    \centering
    \includegraphics[width = 8 cm]{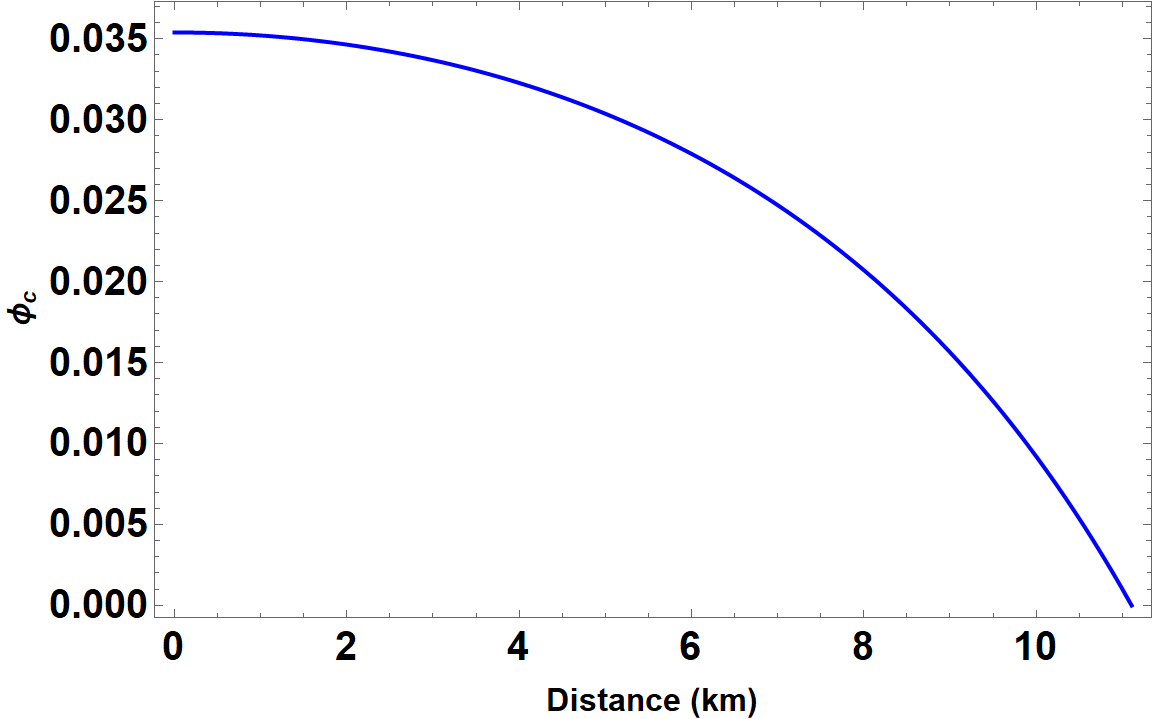}
    \caption{The strength of the scalar field (in units of the Planck mass) vanishes precisely at the surface of the star, just as the radial pressure does.}
    \label{Scalar}
\end{figure}

%%%%%%%%%%%%%%%%%%%%%%%%%

\begin{figure}[h]
    \centering
    \includegraphics[width = 8 cm]{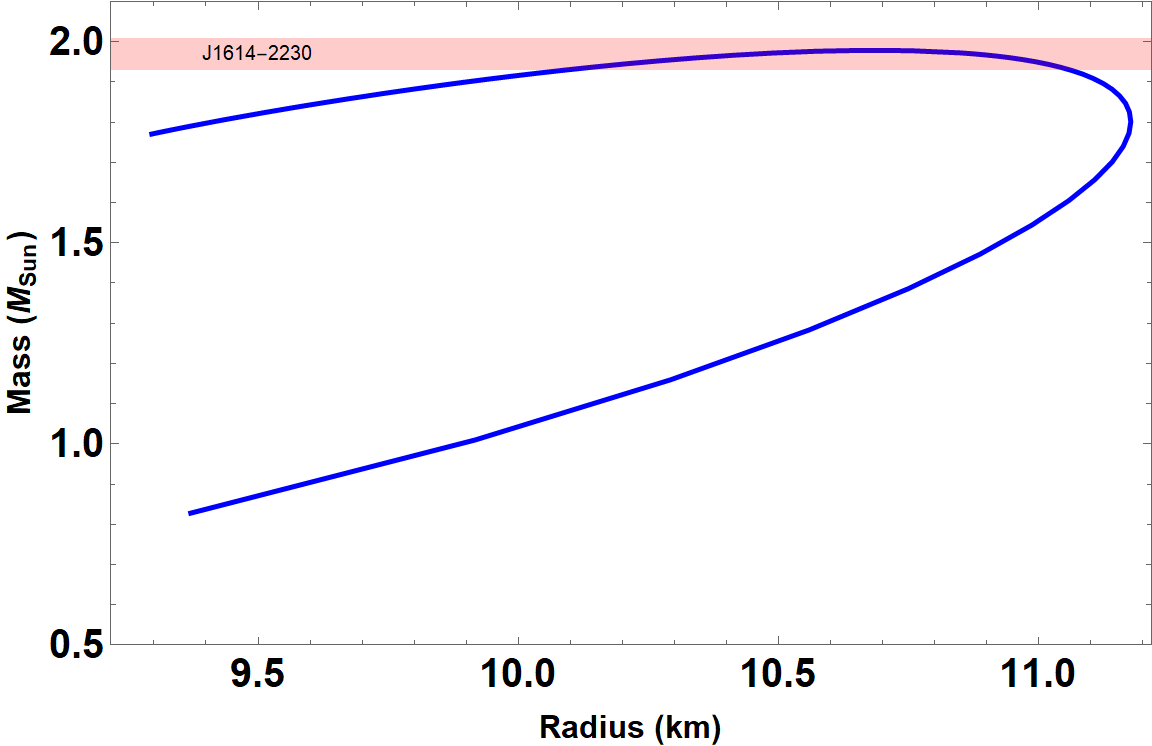}
    \caption{The Mass-radius relation for anisotropic quark stars. The horizontal band shows the observational constraints from pulsar measurements PSR J1614-2230. }
    \label{Profile1}
\end{figure}

%%%%%%%%%%%%%%%%%%%%%

\begin{figure}[h]
    \centering
    \includegraphics[width = 8 cm]{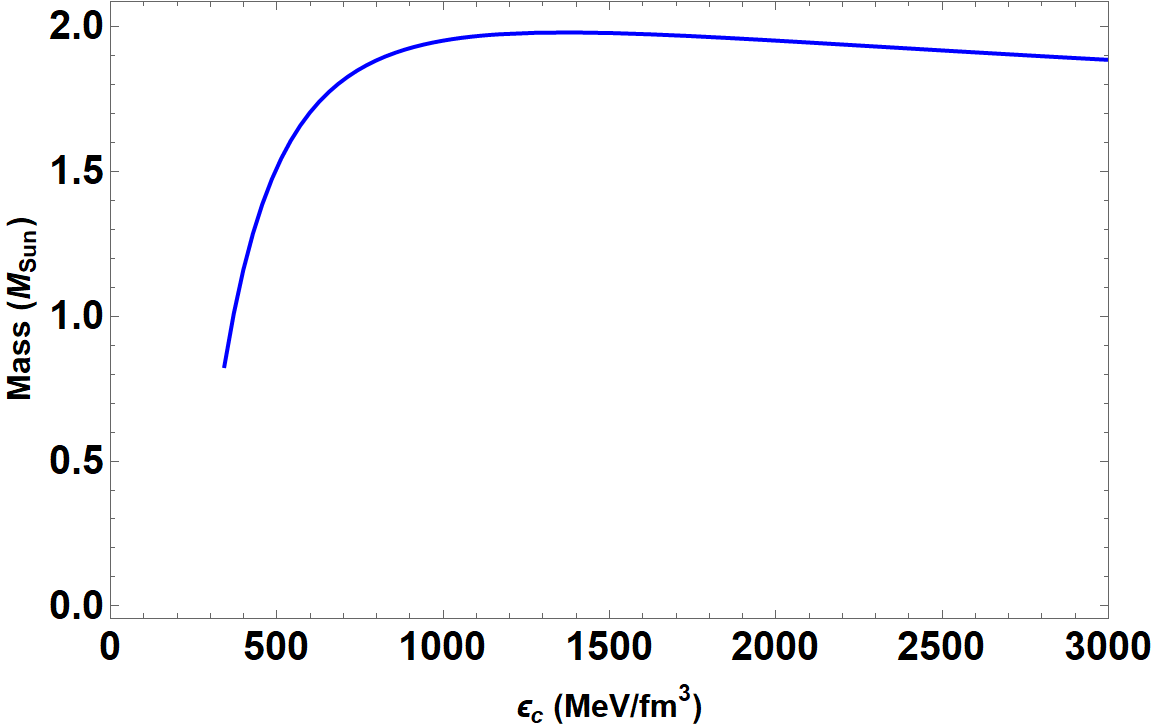}
    \caption{Variation of the total mass versus the central energy density.}
    \label{Stability1}
\end{figure}

%%%%%%%%%%%%%%%%%%%%%%%

\begin{figure}[h]
    \centering
    \includegraphics[width = 8 cm]{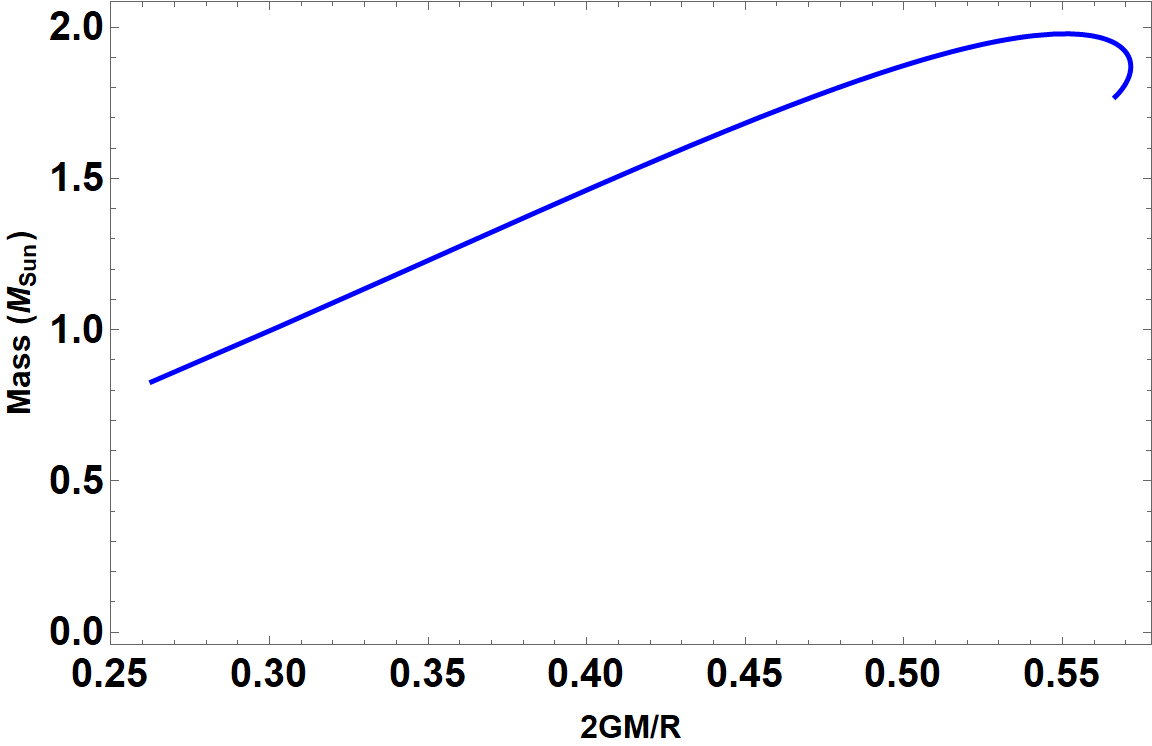}
    \caption{The profile of the maximum mass versus the compactness.}
    \label{compact1}
\end{figure}

%%%%%%%%%%%%%%%%%%%%%%%

\begin{figure}[h]
    \centering
    \includegraphics[width = 8 cm]{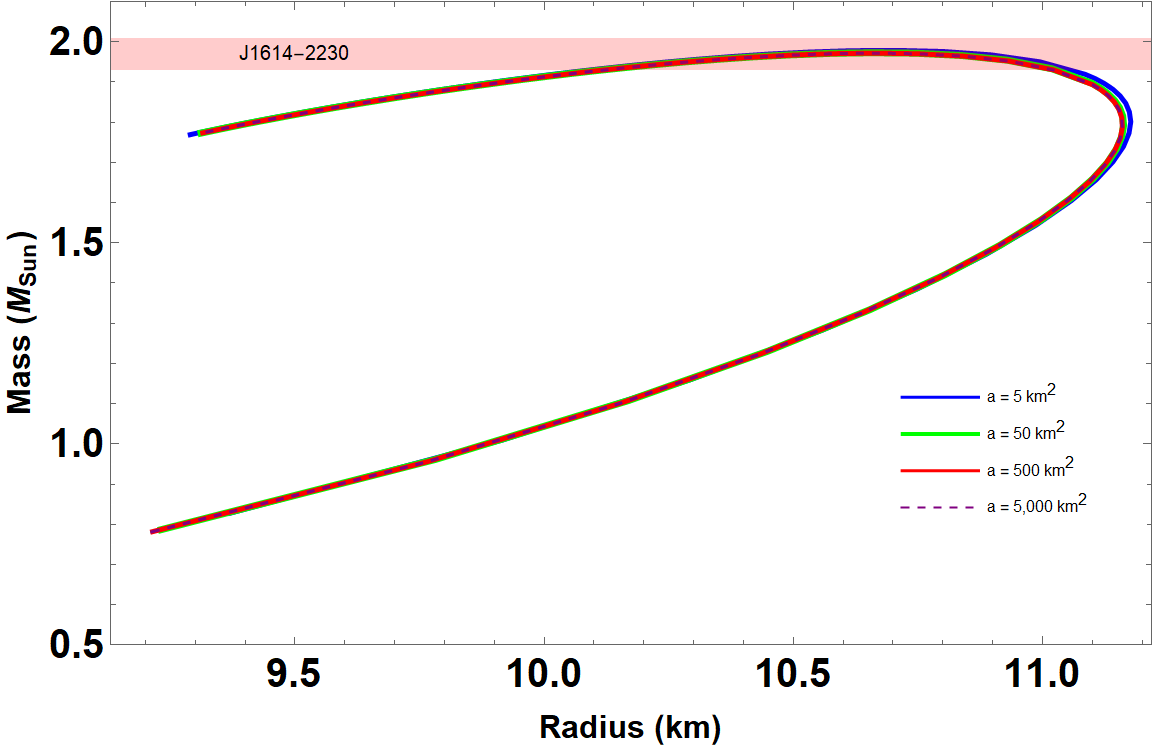}
    \caption{The profile for mass of radius relation.   Different styles and colors of the curves correspond to different values of the parameter $a$. Please see the text for more.}
    \label{Profiles}
\end{figure}

%%%%%%%%%%%%%%%%%%%%%%%%%

\begin{figure}[h]
    \centering
    \includegraphics[width = 8 cm]{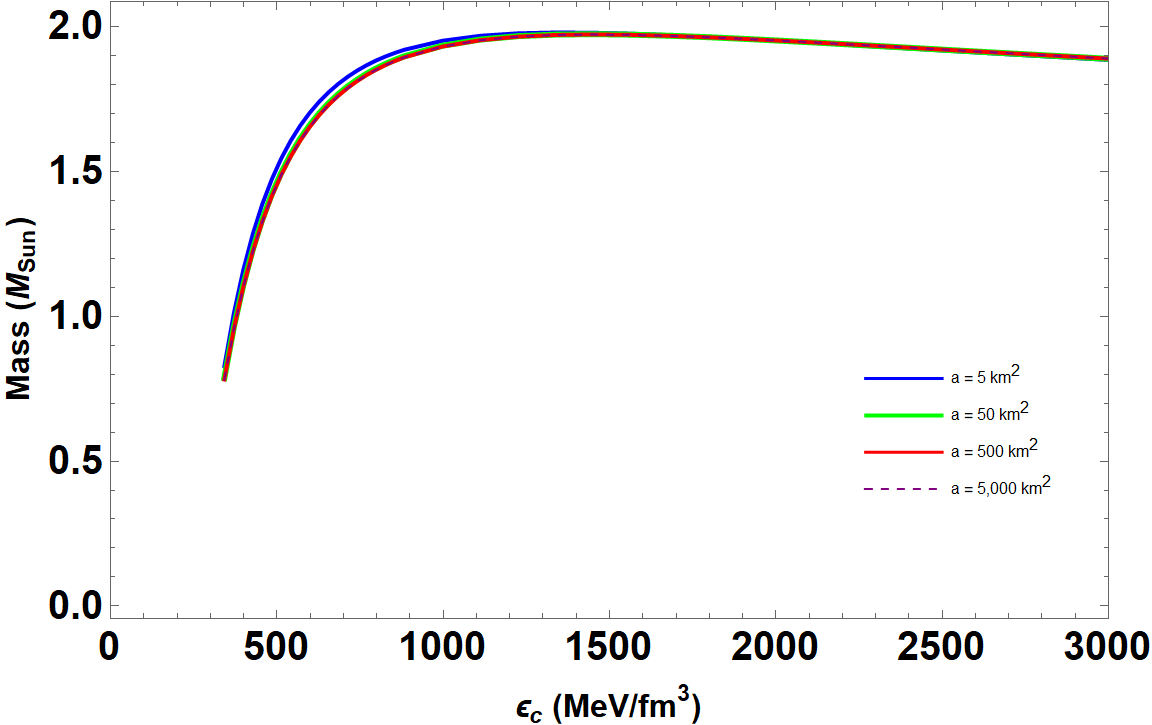}
    \caption{Maximum mass of quark star versus central mass density $\epsilon_c$.}
    \label{Stabilities}
\end{figure}

%%%%%%%%%%%%%%%%%%%%%%%

\begin{figure}[h]
    \centering
    \includegraphics[width = 8 cm]{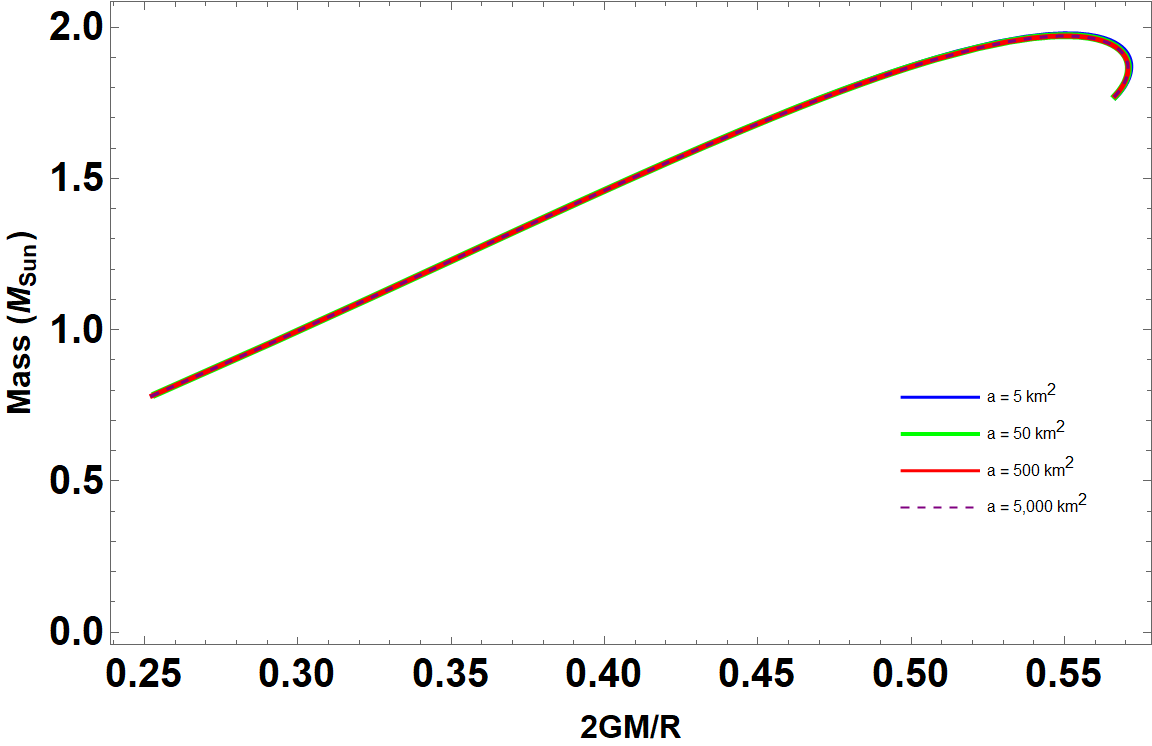}
    \caption{The profile of compactness for different values of the parameter $a$.}
    \label{compact}
\end{figure}

%%%%%%%%%%%%%%%%%%%END-OF-FIGURES%%%%%%%%%%%%%%%%%%%%%%%%%%

%%%%%%%%%%%%%%%%%%%%%%%%%%%%%%%%%%%%
\section{Numerical results}\label{sec4}
%%%%%%%%%%%%%%%%%%%%%%%%%%%%%%%%%%%%

We show the mass-to-radius profile $M-R$, and the behaviour of the basic quantities of interest throughout the star, from the centre at $r=0$ to the surface at $r=R$, assuming $a=5~\text{ km}^2$ first. The radius is measured in $km$, the mass of the stars in solar masses, the energy density and the pressures in $MeV/fm^3$, and the parameter $a$ in $\text{ km}^2$, while the scalar field as well as the factor of compactness are dimensionless quantities.

In particular,  the mass function and the scalar field versus radial coordinate are shown in Figs. \ref{Mass} and \ref{Scalar}, respectively. In Fig. \ref{Pressure} is evident that the tangential pressure is slightly higher than the radial one, and therefore the anisotropic factor is positive. The energy density at the surface acquires its surface value, the radial pressure vanishes, whereas the tangential pressure does not have to vanish at $r = r_S$. This is to be expected in anisotropic stellar models.

Next, we will study some physical properties of neutron star structure. We show the mass of the star versus its radius in Fig. \ref{Profile1} and the mass versus central energy density in Fig. \ref{Stability1}, respectively. In Fig. \ref{compact1} depicts the mass versus the factor of compactness i.e., $C = 2GM/R$. The following features may be observed: i) The profile exhibits the typical behaviour that characterizes quark matter, with a maximum radius first and then a maximum mass.
The mass-radius $(M-R)$ relation depends on the choice of the free parameter $a$ with anisotropic factor
$a_{4} ^\perp = 0.3$, showing the results in Fig. \ref{Profile1}. In the context of our model predicts the existence of the static
 neutron star PSR J1614-2230, whose mass is $M = 1.97 \pm 0.04 M_{\odot}$ \cite{Demorest:2010rz} 
was determined through the Shapiro delay. In the limiting case of $a \to 0$ the solutions
converge to the GR ones, ii) The compactness does not exceed the Buchdahl bound, $C \leq 8/9$ \cite{buchdahl}, and iii) 
to plot Fig.~\ref{Stability1} we take central energy density to be in the range 
$(\sim 360-3000)~MeV/fm^3$. The mass of the star reaches a maximum value at $\epsilon_c^*$,  which is in agreement with what is shown in the $M-R$ profile. What is more, according to the  Harrison-Zeldovich-Novikov criterion \cite{harrison,ZN}, is 
\begin{equation}
\frac{dM}{d \epsilon_c} > 0  \; \; \; \rightarrow \textrm{stable configuration,}
\end{equation}
\begin{equation}
\frac{dM}{d \epsilon_c} < 0  \; \; \; \rightarrow \textrm{unstable configuration.}
\end{equation}
Fig. \ref{Stability1} shows only the first part of the curve, before the maximum value, corresponds to a stable configuration. Therefore, the point at the extremum of the curve separates the stable from the unstable configuration.

In order to be more precise we also made a systematic comparison with the results for different non-negative values for the parameter $a$. Depending on the choice of parameters we investigate the maximum mass, compactness, and the mass versus central energy density (see Figs. \ref{Profiles},  \ref{Stabilities} and \ref{compact}, for more details). Lines with different styles and colors in every figure
correspond to different values of the parameter $a$ ranging from $a = 5$ to $a = 5 \times 10^4~\text{ km}^2$. For positive and progressively increasing values of $a$, one can see that the
$M-R$ diagram is almost indistinguishable from $a = 5$ in Fig. \ref{Stabilities}. Similar trend has been found for 
other curves and maintain all the features observed before.  Furthermore, we observe that  
 the highest mass slightly increases with $a$, which is in agreement with the findings of 
previous related works, see e.g. \cite{Astashenok:2020qds,Astashenok:2021peo,Panotopoulos:2018enj}. Thus, our numerical results 
indicate that the properties of the stars are not sensitive to $a$.

Before concluding our work a comment is in order. In the present article we fixed the numerical values of the parameters that enter into the EoS, as our goal was to investigate the impact of the parameter $a$ on the properties of the stars. In the $M-R$ profiles we have generated the highest mass does not reach the $2.6~M_{\odot}$. In previous related works the authors demonstrated that QCD superconductivity effects imply an EoS for quark matter that can support stars as heavy as $2.6~M_{\odot}$ \cite{Roupas:2020nua,Horvath:2021wmn}, although the authors of \cite{Astashenok:2021peo} concluded that the quark star possibility is not probable.

%%%%%%%%%%%%%%%%%%%%%%%%%%%%%%%%%%%%
\section{ Conclusion }\label{sec5}
%%%%%%%%%%%%%%%%%%%%%%%%%%%%%%%%%%%%

In the present work we have studied relativistic stars in the framework of $f(R)$ theories of gravity. In particular, we have investigated in detail the properties of strange quark stars with anisotropic matter within the well motivated Starobinsky model, characterized by a single parameter $a > 0$. To obtain interior solutions, first we presented the structure equations (generalized TOV equations) describing hydrostatic equilibrium in the Einstein frame, where a canonical scalar field with a certain self-interaction scalar potential is also present. The scalar field is coupled both to gravity and to matter via the trace of the stress-energy tensor of the latter. After that we presented the EoS for quark matter, where the radial pressure and the tangential pressure are certain non-linear functions of the energy density, and there is a non-vanishing anisotropic factor characterizing anisotropic matter. We have integrated numerically the structure equations, and  matched the interior solution to the exterior vacuum solution corresponding to the usual Schwarzschild geometry. We have made sure that both the radial pressure and the scalar field vanish simultaneously at the surface of the star, and we have computed the radius, the mass as well as the factor of compactness of the star. The behaviour of the solution for the mass function, the energy density, the radial and tangential pressure as well as the anisotropic factor have been shown graphically for $a = 5 \text{ km}^2$ and $\epsilon_c = 10~B$. For the same value of $a$ and varying the central value of the energy density, we have obtained the mass of the star as a function of the radius of the star, its factor of compactness and the central energy density. The highest mass of the star that the assumed EoS can support is shown in all three figures. An agreement may be achieved with super-massive observed pulsars at two solar masses. Finally, to see the impact of the parameter $a$ on the properties of the star, we have computed the $M-R$ relationships considering increasing values of $a$ from $5~\text{ km}^2$ to $5 \times 10^3~\text{ km}^2$. Our numerical results show that the obtained curves are only slightly shifted compared to the original ones corresponding to $a = 5~\text{ km}^2$. The same holds for the other two cases, namely mass of the star versus central energy density, and mass of the star versus factor of compactness. 
We thus conclude that the properties of the stars are very little sensitive to the parameter $a$.

%%%%%%%%%%%%%%%%%%%%%%%%%%%%%%%
\section*{Acknowledgements}

 We wish to thank the anonymous reviewer for useful suggestions.
The author G.~P. thanks the Fun\-da\c c\~ao para a Ci\^encia e Tecnologia (FCT), Portugal, for the financial support to the Center for Astrophysics and Gravitation-CENTRA, Instituto Superior T\'ecnico, Universidade de Lisboa, through the Project No.~UIDB/00099/2020 and No.~PTDC/FIS-AST/28920/2017. 

%%%%%%%%%%%%%%%%%%%%%%%%%%%%%55

\end{document}